\documentclass[preprint,showpacs,preprintnumbers,amsmath,amssymb,floatfix]{revtex4}

\usepackage{epsfig}
\usepackage{dcolumn}
\usepackage{bm}

\begin{document}

\preprint{APS/123-QED}

\title{On the intermittent character of \\ interplanetary magnetic field fluctuations}

\author{Roberto Bruno}
\affiliation{IFSI/INAF - via Fosso del Cavaliere, Roma - Italy}
\author{Vincenzo Carbone}
\affiliation{Dipartimento di Fisica - Universit\`a della Calabria, and CNISM, Unit\`a di Cosenza, Arcavacata di Rende (CS) - Italy}
\author{Sandra Chapman}
\author{Bogdan Hnat}
\affiliation{Centre for Fusion, Space and Astrophysics, University
of Warwick, UK}
\author{Alain Noullez}
\affiliation{Observatoire de la C\^ote d'Azur, Bd. de l'Observatoire, Nice - France}
\author{Luca Sorriso-Valvo}
\affiliation{LICRYL - INFM/CNR, Arcavacata di Rende (CS) - Italy}
\email{sorriso@fis.unical.it}

\date{\today}

\begin{abstract}
Interplanetary magnetic field magnitude fluctuations are
notoriously more intermittent than velocity fluctuations in both
fast and slow wind. This behaviour has been interpreted in terms
of the anomalous scaling observed in passive scalars in fully
developed hydrodynamic turbulence. In this paper, the strong
intermittent nature of the interplanetary magnetic field is briefly discussed
comparing results performed during different phases of the solar cycle. 
The scaling properties of the interplanetary magnetic
field magnitude show solar cycle variation that can be
distinguished in the scaling exponents revealed by structure
functions. The scaling exponents observed around solar maximum
coincide, within the errors, to those measured for passive scalars
in hydrodynamic turbulence. However, it is also found that the values
are not universal in the sense that the solar cycle variation may
be reflected in dependence on the structure of the velocity
field.
\end{abstract}

\pacs{94.05.-a, 96.50.-Bh, 96.50.-e, 47.65.-d, 47.27.-i, 47.27.-Jv, }
\maketitle

\section{\label{sec:level1}Introduction}

The interplanetary medium is pervaded by fluctuations providing
information on plasma turbulence on a large range of scales, from
fractions of second up to the solar rotation period
\cite{tuemarsch}. Some characteristic features of
these fluctuations include \cite{living} strong anisotropy shown by
velocity and magnetic field fluctuations, the different radial
evolution of the minimum variance direction of the magnetic field
(hereafter IMF) and velocity, the lack of equipartition between
magnetic and kinetic fluctuations as well as scaling and non- Gaussian
Probability Density Functions (PDF) of flux densities \cite{hnatpre}.
All those factors contribute to the view that statistical properties of
the solar wind should be far from that of Kolmogorov \cite{k41} for
fluid turbulence which assumes isotropic, homogeneous and
incompressible flow.

However, as shown since the first in situ observations
\cite{coleman}, fluctuations measured in the solar wind plasma
parameters share many statistical features with fluctuations
observed in hydrodynamic turbulence~\cite{tuemarsch,living}. For
example, the problem of intermittency, namely the departure from
statistical self-similarity, has been investigated on the basis of
standard techniques and modelling~\cite{living}. The ``strength"
of intermittency, related to the ability of the turbulent cascade
to develop singularities at small scales, can be quantified by
measuring the scaling exponents $\zeta_p$. These are defined
through the structure functions  $S_p(\tau)=\langle (\delta
v_\tau)^p \rangle \sim \tau^{\zeta_p}$, where $\delta
v_\tau(t)=v(t+\tau)-v(t)$ are the velocity fluctuations over a
given scale $\tau$~\cite{frisch,burlaga91,living}. Similarly, we
can define structure functions, and exponents, for fluctuations in
temperature, magnetic field magnitude or density in order to investigate
their scaling properties.

The scaling exponents for velocity and magnetic
field~\cite{hnat,burlaga91,carbone95} are nonlinear in $p$,
departing from the Kolmogorov non-intermittent scaling $\zeta_p =
p/3$~\cite{k41,frisch}. The scale dependent character of the field
fluctuations can also be observed from the shape of the PDFs of the
fields~\cite{marsch&liu93,sorriso99}. In this case the PDFs of
fluctuations standardized with their relative standard deviation
increasingly depart from a Gaussian distribution with decreasing
temporal scale $\tau$. Recent studies have shown that the slow
wind is more intermittent than the fast wind and, as largely
reported in literature, the solar wind magnetic field is more
intermittent than the velocity
field~\cite{marsch&liu93,carbone95,sorriso99,sorriso04,living}.
The difference between velocity and magnetic field intermittency 
has also been observed in two-dimensional incompressive 
magnetohydrodynamics (MHD) numerical simulations~\cite{popoca,sorriso00,sorriso01,merri}, 
and seems to be a robust feature of MHD turbulence.
Such behaviour is similar to the transport of passive fields in
fully developed hydrodynamic turbulence where the advected field
(usually temperature) is more intermittent than the advecting
velocity field~\cite{vergassola}. This suggests a similarity
between statistical features of the interplanetary magnetic field
strength and hydrodynamic passive scalars~\cite{sorriso99,sorriso04,living}. 
The question then arises as to whether these similarities are a consequence 
of shared dynamical behaviour. Indeed, the magnetic field plays a
relevant role in both velocity shear instability and parametric
instability during the development of the turbulence spectrum
observed in the solar wind~\cite{malara,bavassano2000}.

\begin{figure}
\centering{\includegraphics[width=10cm]{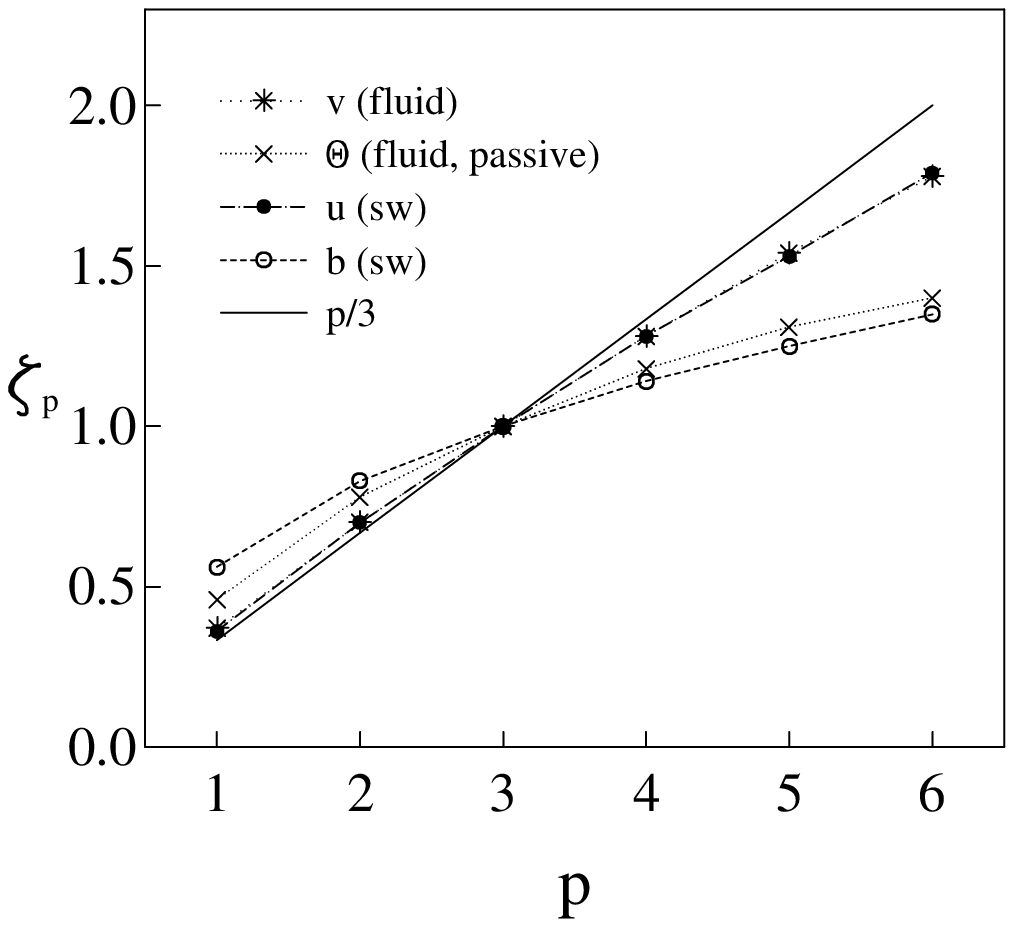}} 
\caption{} 
\label{fig:fig01}
\end{figure}

Figure~\ref{fig:fig01} shows the behaviour of scaling exponents
$\zeta_p$ versus order $p$, computed through the Extended
Self-Similarity technique~\cite{ess} and normalized to the third
order scaling exponent $\zeta_3$ for both solar wind velocity and
magnetic field, as compared with results from passive scalar
transported by incompressible hydrodynamic turbulence. The
interplanetary data were obtained during a slow wind interval
observed at the Helios 2 satellite at 0.9 AU while the fluid data
derive from the experiment~\cite{benzi}. We can see from the plot
that the exponents for the magnitude of velocity for both fluid
and solar wind approximately coincide, as do those of the solar
wind   magnetic field magnitude and the fluid passive scalar. The
question then is whether or not such  behaviour can be ascribed to
the passive nature of the interplanetary magnetic field and as such
is universal.

Passive scalar dynamics for the interplanetary magnetic field
(IMF) magnitude would have  far reaching consequences for the
theoretical description and the modelling of solar wind
turbulence.
For example, if the magnetic field could be treated as a passive
vector the observed Kolmogorov-like spectra of the solar wind
\cite{tuemarsch} could be explained using hydrodynamic turbulence.
In this paper, we test whether this 'passive scalar' signature in
the IMF is robust by comparing scaling properties of the IMF
magnitude at solar minimum and maximum derived from ACE and WIND
spacecraft data sets. Previous studies of the energy input that the
solar wind provides to the magnetosphere have shown that a measure of
the solar wind Poynting flux shows scaling with exponents that
vary with the solar cycle \cite{hnatjgr}.

\section{Data Analysis}

We will now test whether or not the passive scalar
characteristics of interplanetary magnetic field intensity
fluctuations are found consistently during different solar cycle
phases. If the IMF magnitude is intrinsically a passive scalar,
its scaling should be robust against the particular data sample
that we choose, provided that the fluctuations are in a state of fully
developed turbulence. We will first perform an analysis of the
data sets corresponding to extended intervals around solar minimum
and maximum. We utilise two data sets, from the WIND and ACE
spacecraft. The $92$ second average WIND spacecraft data
spanning a single year $1996$ will provide our solar minimum
sample, whilst the ACE spacecraft $64$ second average data from
the year $2000$ will provide our  solar maximum sample.

\begin{figure}[t]
\epsfsize=0.475\textwidth
\leavevmode\epsffile{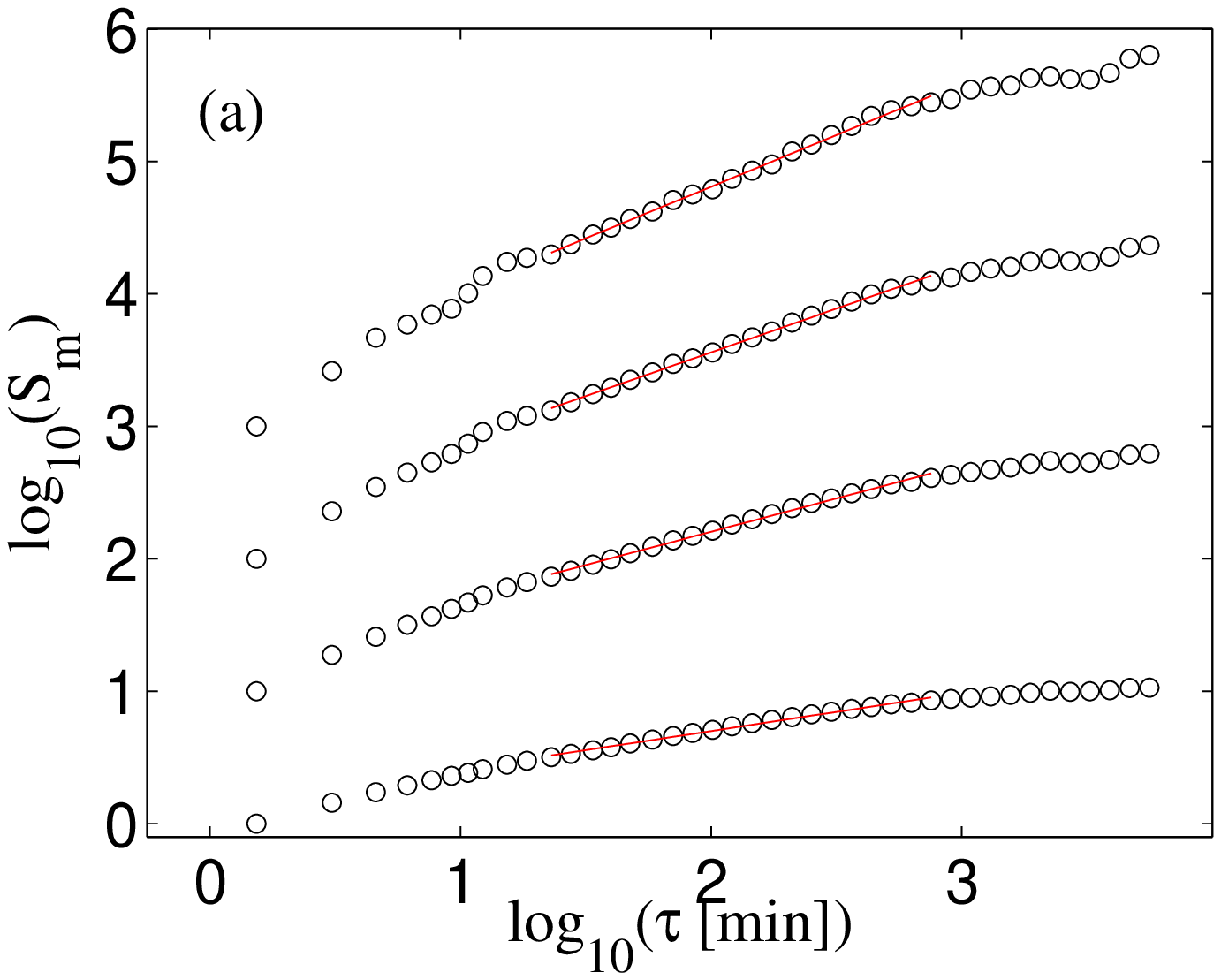}
\epsfsize=0.475\textwidth
\leavevmode\epsffile{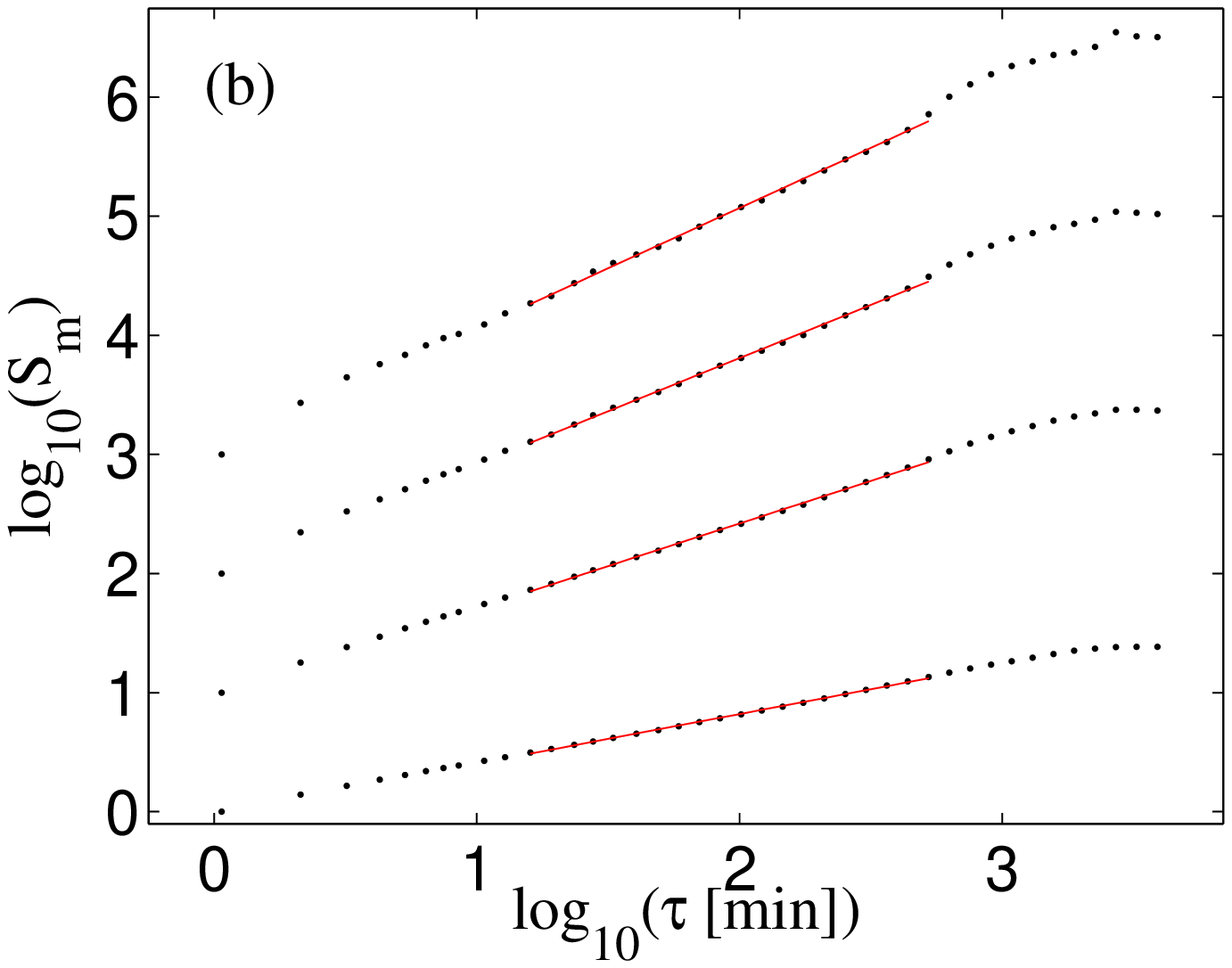}
\caption{}
\label{fig:fig02}
\end{figure}

\begin{figure}[t]
\epsfsize=0.475\textwidth
\leavevmode\epsffile{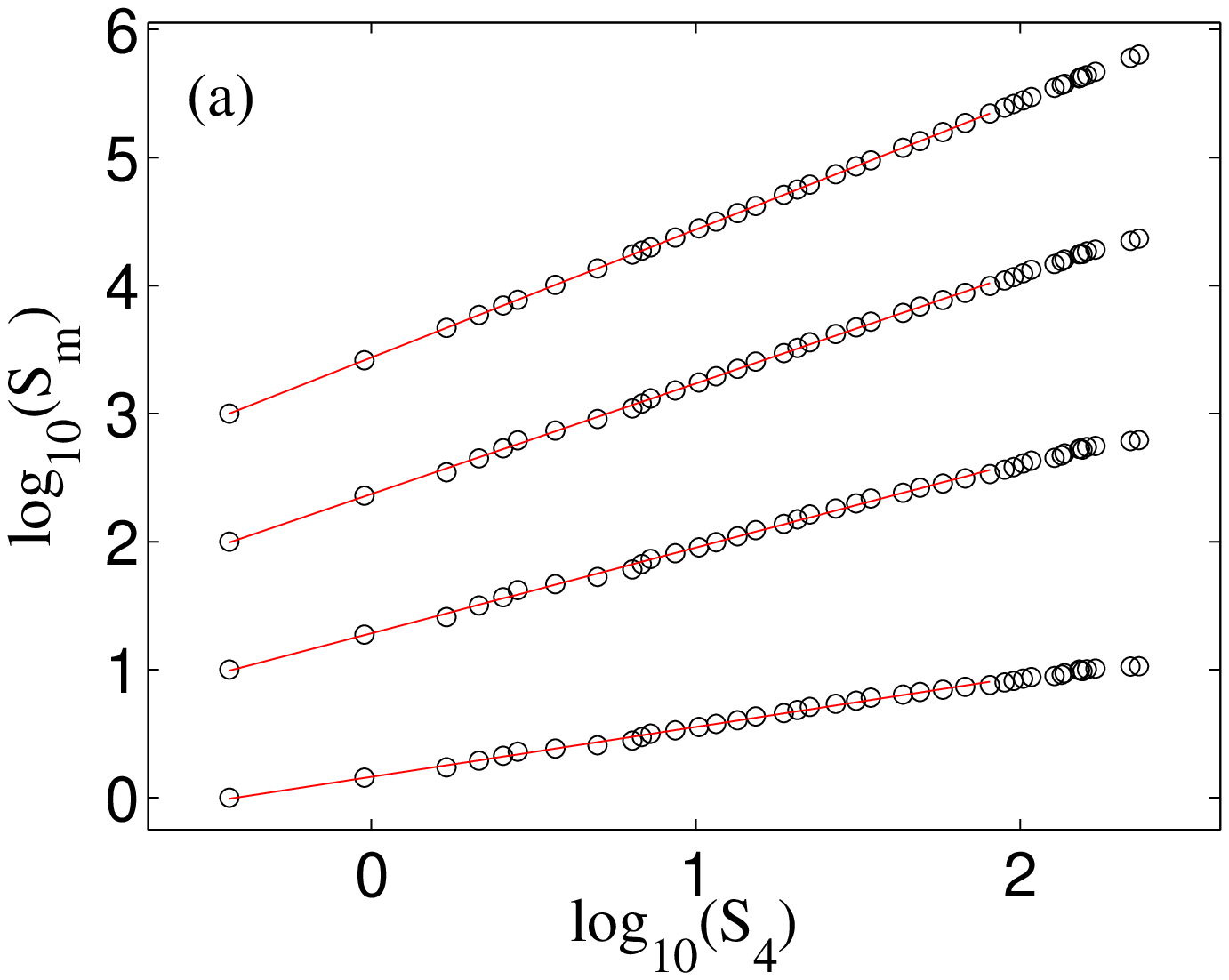}
\epsfsize=0.475\textwidth
\leavevmode\epsffile{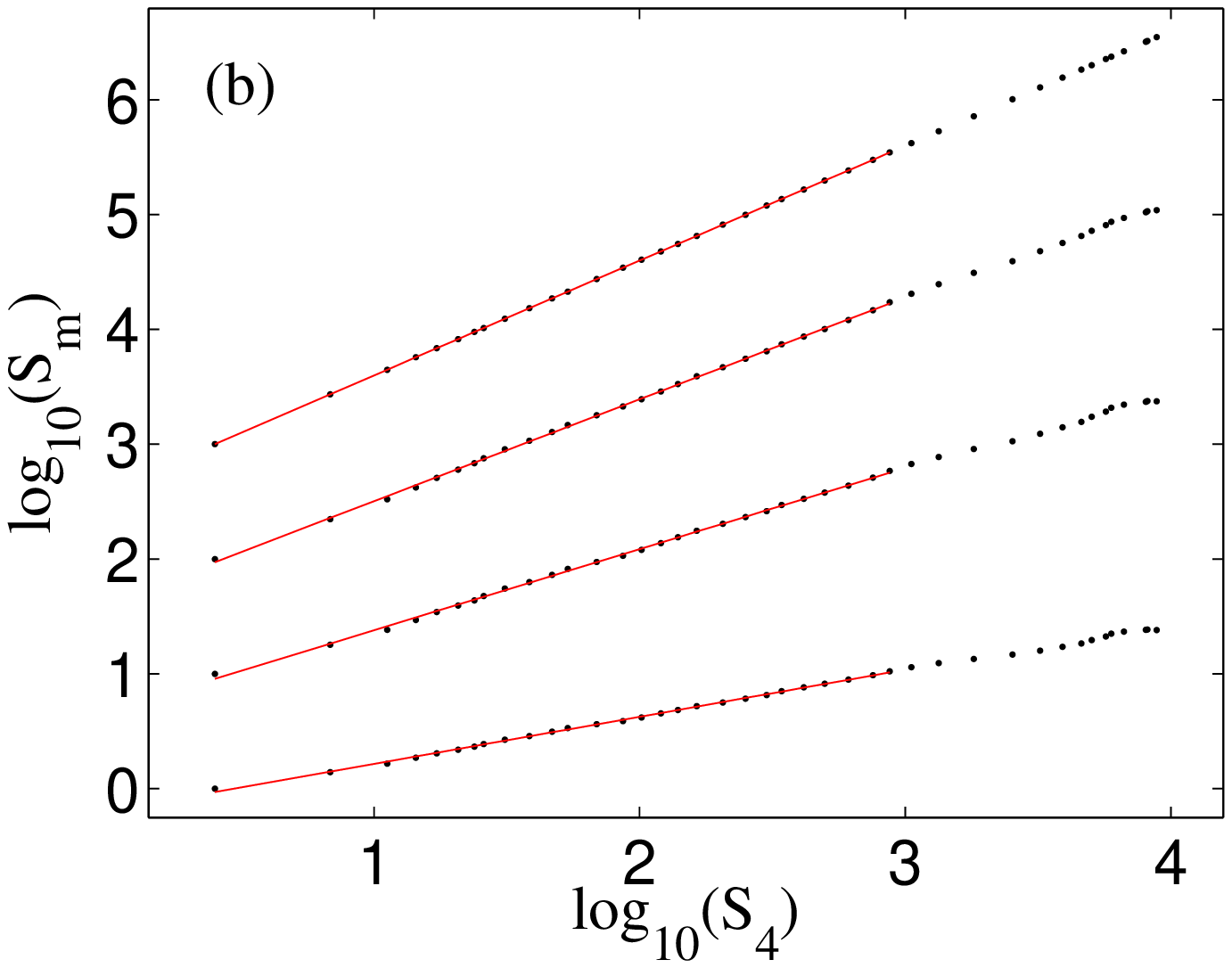}
\caption{}
\label{fig:fig03}
\end{figure}

Figure~\ref{fig:fig02} shows the structure functions of magnetic
field magnitude for solar minimum and maximum respectively (offset
vertically for clarity). The common scaling region from which
these structure functions are obtained are shown by the solid
lines on the Figure~\ref{fig:fig02}, these represent the best
regression fits for temporal scales $\tau$ between $\sim 10$
minutes and $\sim 10$ hours. These scaling regions can be
significantly extended, for both data sets, by means of Extended
Self-Similarity (ESS). The ESS method seeks scaling
$S_m(\tau)\!\propto\! S_p^{\eta(m)}(\tau)$ which should emerge on
a plot of $S_m$ versus $S_p$. We plot $S_m$ versus $S_4$ on
logarithmic axes for fluctuations in $B$, for solar minimum and
maximum in Figure~\ref{fig:fig03} (a,b) respectively. We use this
extended scaling range to obtain an improved estimate of the
exponents $\tilde{\zeta}(m)\!=\!\zeta(4) \eta(m)$, where
$\zeta(4)=0.78 \pm 0.03$ for solar minimum and $\zeta(4)=1.01 \pm
0.03$ for solar maximum.

\begin{figure}[t]
\epsfsize=0.75\textwidth
\leavevmode\epsffile{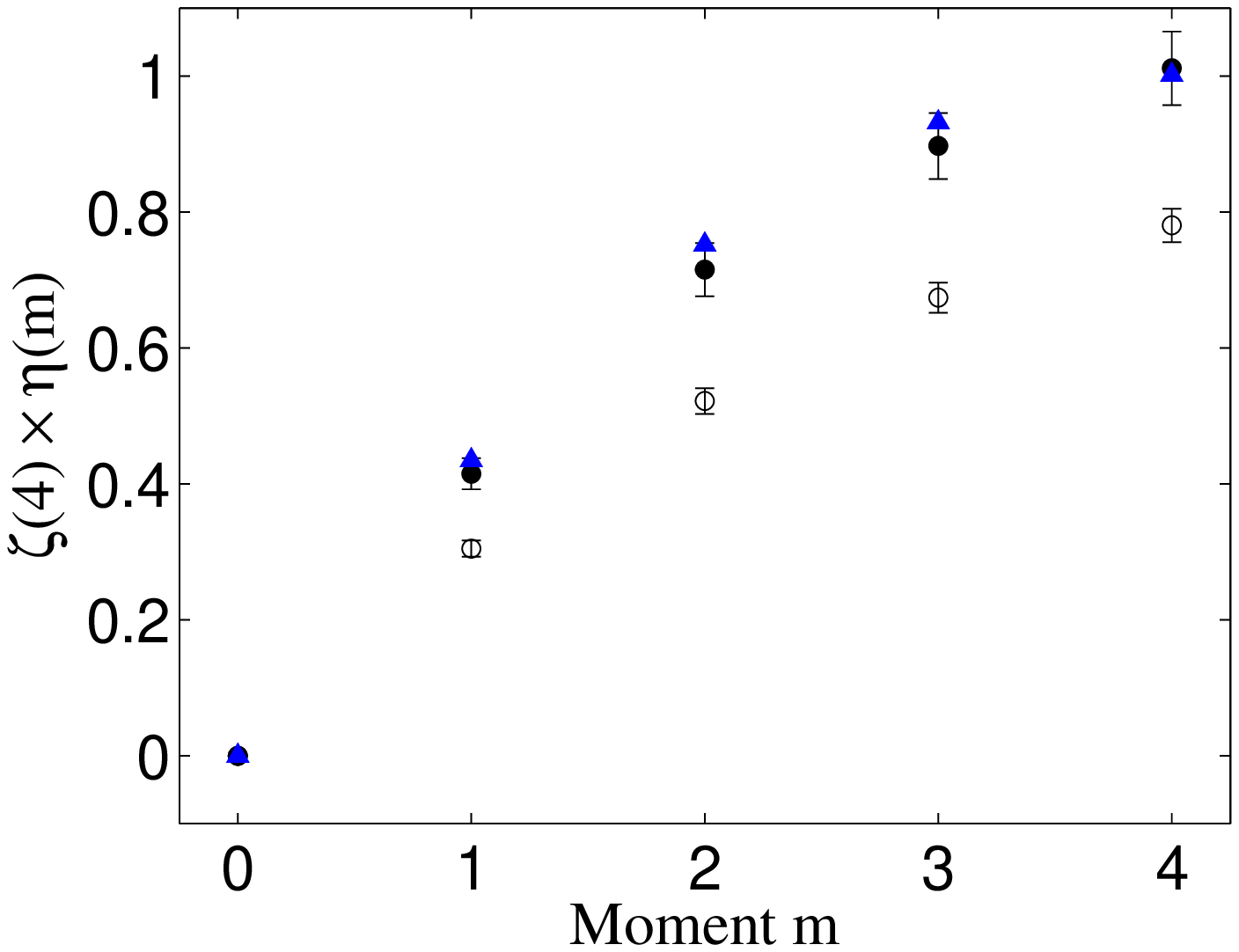}
\caption{}
\label{fig:fig04}
\end{figure}

Figure~\ref{fig:fig04} combines scaling exponents $\zeta(4)
\eta(m)$ obtained from ESS for the magnetic field intensity
fluctuations during solar minimum and maximum. For comparison, we
have also plotted scaling exponents obtained from the data set
used in Ref.~\cite{papero}. The result clearly demonstrates that
the magnetic field magnitude fluctuations do not exhibit a
single and universal scaling that coincides with that found in the
fluctuations of passive scalars in hydrodynamic turbulence. It is
also clear, however, that there is a very good agreement between
scaling exponents found in Ref.~\cite{papero} and these derived
here from the data set corresponding to solar maximum. Recently,
it has been suggested that the large and statistically
under-represented events can obscure the underlying statistical
properties of turbulent data sets~\cite{chapmanNPG05}. A
conditioning method has been proposed to filter out a limited
number of data points and obtain robust scaling properties of the
remaining data. We have verified that such conditioning does not
change the qualitative difference between solar minimum and
maximum, nor the individual scaling exponents which vary within
the error bars. This is qualitatively very different from for
example the behaviour of the density fluctuations which exhibit
substantial differences between non-filtered and filtered data
sets~\cite{hnat}.

To gain a better appreciation of the different scaling regimes of
the IMF magnitude under changing velocity field we will now focus
on different short time intervals with pronounced, well
defined structures. Examining these short intervals has the
advantage over the study of extended intervals in that we
can identify the morphology of individual sequences of data with
particular scaling behaviour. However, the disadvantage is that
these shorter intervals necessarily reduces the number of data
points available, thus we cannot repeat the full structure
function analysis above.

\begin{figure}
\centering{\includegraphics[width=10cm]{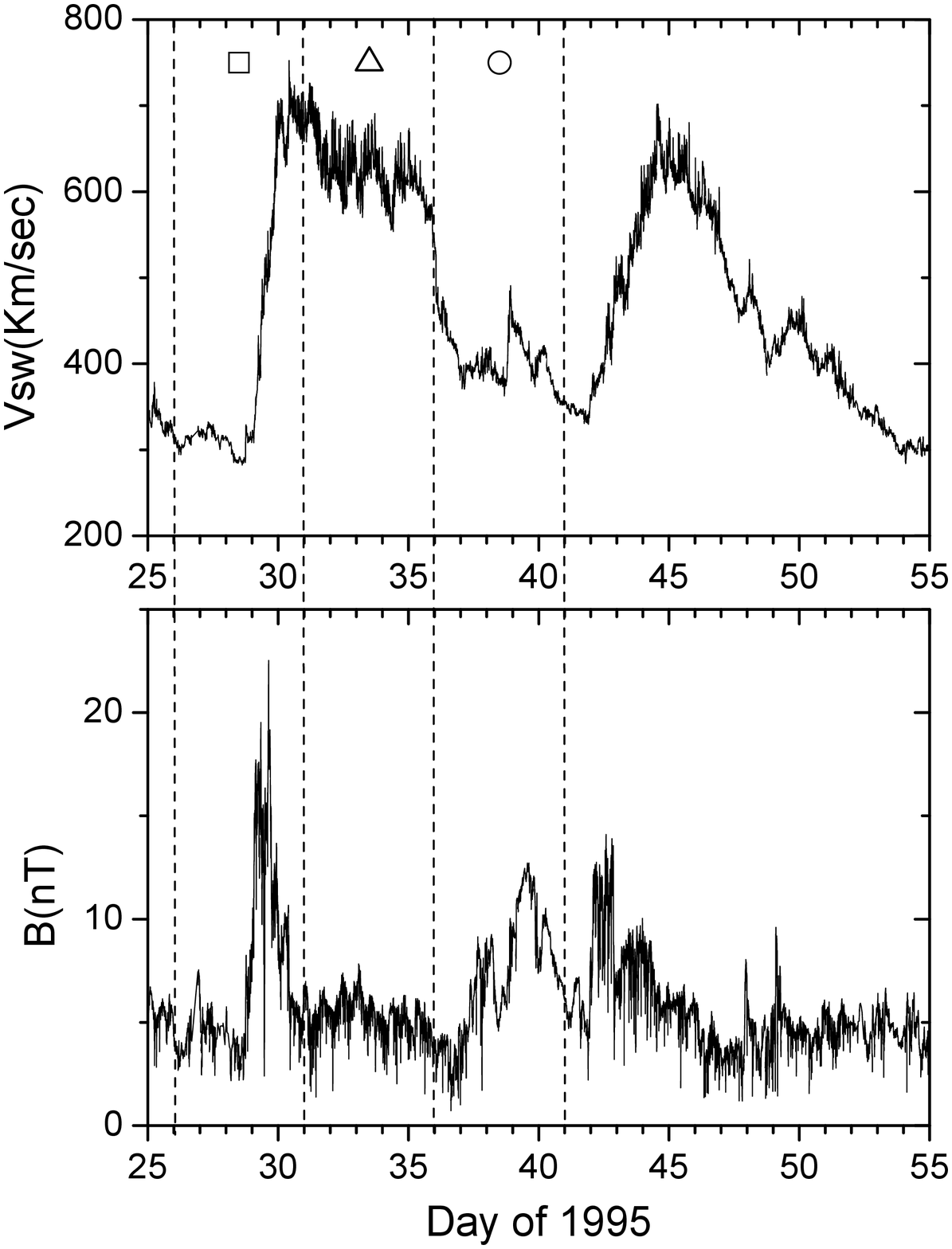}} 
\caption{} 
\label{fig:fig05}
\end{figure}

\begin{figure}
\centering{\includegraphics[width=10cm]{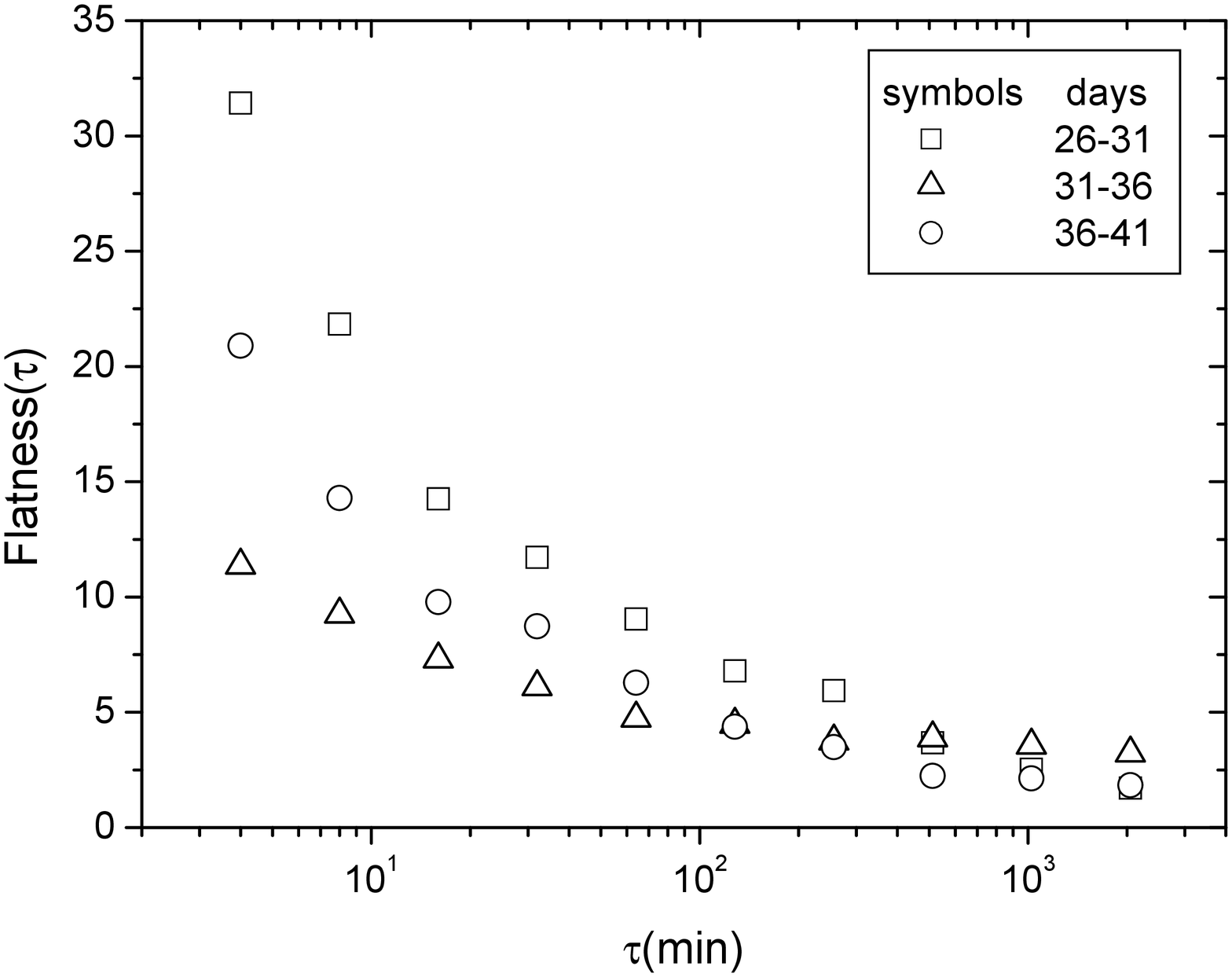}} 
\caption{} 
\label{fig:fig06}
\end{figure}

The top panel of Figure~\ref{fig:fig05} shows the wind speed
profile during $30$ days of data recorded by WIND when the
spacecraft passed through two corotating high velocity streams
beginning on day $30$ and day $43$, respectively. The bottom panel
highlights the different plasma regimes explored by the magnetic
field intensity. This parameter is highly compressed at the stream
interface, especially where the interface is more developed, and
much less within the trailing edge of the stream. Moreover, the
low velocity wind (between days 36 and 42) shows the typical
compressive region of the interplanetary current sheet (see
references cited in~\cite{living}). The four vertical dashed lines
and the three different symbols identify three different regions
where we quantify the intermittent character of the fluctuations
by looking at the flatness factor of the different probability
distribution functions at different scales 
(see Table~\ref{tab1} for data description).
Following~\cite{frisch}, a random function is intermittent at
small scales if the flatness of its fluctuations grows without
bound at smaller and smaller scales. Figure~\ref{fig:fig06} shows
that the three different regions show different values of the
flatness, and suggest that the most compressive region, which display the
highest $rms$ values of the IMF (see Table~\ref{tab1}), is also the most
intermittent region.

The information obtained from the statistics of extended
intervals, where the solar wind morphology is not specified, and
from ``hand selected" shorter intervals of known morphology, is
complementary but not necessarily coincident. Extended studies
will tend to be dominated by the scaling properties of the bulk of
the data (see, for example \cite{hnatpop}). In this context the
presence of large scale coherent   structures will tend to destroy
the scaling, this effect can be removed by conditioning the data
(removing isolated outliers as above). To maintain good statistics
however, the extended interval should clearly not be dominated by
large scale structures such as stream-stream interfaces, Coronal Mass
Ejections and interplanetary shocks which are distinct from the
evolving turbulence. Short intervals, selected by morphology,
explicitly remove these large scale coherent structures. The
dynamical interactions experienced by the solar wind during its
expansion~\cite{tuemarsch} would be expected to deviate from the
ideal condition of incompressibility particularly in the vicinity
of these structures.

\begin{table} [h]
\begin{center}
\caption{The periods of WIND measurements used in this work: 
average and root mean square velocity and IMF, and the size of each dataset.}
\label{tab1}
\begin{tabular}{cccccc}

\hline
days of 1996 & $\langle V\rangle (Km/s)$ & $V_{rms} (nT)$ & $\langle B\rangle (Km/s))$ & $B_{rms} (nT)$ & data points  \\
\hline
$26$---$31$ & $496$   & $160$  & $7.22$   & $4.22$  & $1495$   \\
$31$---$36$ & $622$   & $34$   & $5.31$   & $0.89$  & $1772$   \\
$36$---$41$ & $407$   & $32$   & $7.07$   & $2.83$  & $1798$   \\

\hline
\hline

\end{tabular}
\end{center}
\end{table}

\section{Comparison with other studies}

It is worthwhile to discuss these results in the context of the
Ref.~\cite{papero} where the authors performed similar analysis on
a data sample recorded by ACE during $1998$, that is close to
solar minimum. The values of scaling exponents, however, were
close to these obtained here for solar maximum and it was
concluded, showed passive scalar behaviour. How can we explain
this contradiction? The year $1998$, although not coincident with
the solar maximum of activity cycle $23$, was already
characterized by an enhanced coronal activity if compared to the
preceding solar minimum, about two years before~\cite{Gopalswamy}.
Contrary to what has been stated in Ref.~\cite{papero}, the sun
was certainly not quiet at all during this period, and as reported
by Ref.~\cite{Gopalswamy}, the CME daily rate was about $4.5$
compared to a rate of approximately $0.5$ around the preceding
solar activity minimum which can be considered a truly quiet
period. However, different phases of the solar activity cycle are
also distinguishable because of a completely different
organization of the large scale IMF (see the review~\cite{living}
and references therein).

It was also proposed in Ref.~\cite{papero}  that the necessary
conditions for the magnetic field magnitude evolution equation to
be in the form of advection-diffusion equation. The generic
evolution equation takes form:
\begin{equation}
\partial_t B = -({\bf u} \cdot \nabla)B + \eta \nabla^2 B + \lambda B\ ,
\label{eq:induction}
\end{equation}
where $\bm u$ is an incompressible velocity field, $\eta$ is the
magnetic diffusivity and $\lambda$ is a pseudo dissipation coefficient
given by
\begin{equation}
\lambda = n_\alpha n_\beta\,\partial_\alpha v_\beta
  - \eta\,\partial_\alpha n_\beta \,\partial_\alpha n_\beta\ ,
\label{eq:lambda}
\end{equation}
while the vector $\bm {n}=\bm {B}/B$ is the unit vector in the
magnetic field direction. The similar scaling observed for the
magnetic field magnitude and the passive scalar advected by
hydrodynamic turbulence suggests conditions in which the $\lambda$
stretching term could vanish, in which case the evolution of the
magnetic field magnitude would reduce to that of a passive
scalar~\cite{papero}.
However in~\cite{papero}  only  conditions in which the first term
of~(\ref{eq:lambda}) would vanish were considered, with the
assumption that the second term must be small since the magnetic
diffusivity ~$\eta$ must be small.  While the precise value
of~$\eta$ in the solar wind is unknown,  it must indeed be small
since the magnetic Reynolds number ${\rm Re}_{\rm m} = UL/\eta$ is
observed to be large \cite{matt05}; however it does not
necessarily follow that the second term of~(\ref{eq:lambda}) is
small. In turbulence at large Reynolds number the gradients of the
field or the gradients of the field directions (because the field
value themselves are bounded by a few times~$B_{\rm rms}$) can
become very large, in order to maintain a finite limit of the
dissipation when the diffusivity~$\eta$ goes to zero. Such
``extreme'' gradients have been observed (\cite{bobbybrown2}), in
which the magnetic field changes direction in an abrupt fashion,
while keeping its magnitude unchanged. In these cases, it is
precisely this second term which is responsible for
preserving~$B$, because it then acts as a damping term which
cannot be neglected.  This second term in addition has a well
defined (negative) sign and thus cannot vanish by averaging.

In order for (\ref{eq:induction}) to be similar to a passive
scalar evolution equation, the first term of~(\ref{eq:lambda})
should also vanish and some conditions in which this could be
realised are given in~\cite{papero}.  However these conditions are
{\em not\/} realised most of the time, and not in the fast solar
wind~\cite{living}. The conditions are essentially
{\em loss of correlation\/} between the magnetic field direction and
of the velocity gradient, however this will not be the case in
Alfv\'enic turbulence  in the fast wind, where the
velocity and magnetic fields follow each other closely,
anticorrelated (for outward Alfv\'en waves) or correlated (for
inward waves) fashion \cite{living}. In these cases, the magnitudes of
$\bm B$ and~$\bm u$ also follow each other closely, and it is clear
that~$B$ cannot be considered as a passive scalar. The first term
of~(\ref{eq:lambda}) then also  ensures the reaction of the
magnetic field on the velocity and the equipartition between
magnetic and kinetic energy. Recent studies also indicate that the
assumption of incompressibility of the plasma is not consistent
with passive scalar behaviour in the solar wind~\cite{hnat}.

\section{Conclusions}
The aim of this study was to gain a better understanding of the
scaling properties of the IMF magnitude and its dependence on the
changing solar wind velocity field. Our results show a remarkable
similarity between the scaling exponents of  passive scalars in
hydrodynamic turbulence and the IMF magnitude during solar
maximum. At $\approx 1$ AU, slow streams that originate from the
near-equatorial region of the Sun, dominate the solar wind at
solar maximum. This result is in agreement with our Figure $1$,
where scaling properties of slow solar wind streams were
presented. We draw the readers attention to the fact that the
scaling of the velocity fields for the hydrodynamic flow and the
solar wind also exhibit nearly identical scaling in that case.
Statistical properties of the passive fluctuations are known to
depend strongly on the characteristics of the advecting velocity
field~\cite{passive_anz}. In that respect it is not surprising
that the scaling properties of the IMF magnitude change
dramatically with the solar cycle.

Further work is needed to establish how universal these scaling
properties of the IMF magnitude really are. This could be done by
applying conditional statistics, that is deriving the scaling
properties of $B$ from the data sets with similar velocity
profiles. If scaling exponents derived from these intervals are
robust we could then accept the scaling as universal. This is,
however, a rather daunting task, as illustrated by Figure $5$. The
analysis requires long data sets in order to obtain the scaling
exponents with sufficient precision, however
 the velocity profile of the solar wind is
very dynamic, leaving us with relatively short time intervals.

We now arrive to the most intriguing question: Is the magnetic field
magnitude passively advected in the solar wind? It is clear that one
can draw only partial conclusions from our results. We have shown that
there is some evidence, based on scaling properties of the IMF magnitude,
that the passive scalar model for $|B|$ could, indeed, be valid for
the slow solar wind. The universality of the scaling is, however, still to
be addressed. This is even more so for the fast solar wind where the departure
from the known hydrodynamic scaling is very pronounced.

We also point out the apparent difficulty in relating the known features of
the slow and fast streams of the solar wind to the assumptions of Ref
\cite{papero} which led to an evolution equation for $|B|$ written as an
advection-diffusion one.
The assumptions used for this were: (i) incompressibility of the flow
($\nabla \cdot \bm{u}=0$), and (ii) the isotropy of the average direction of
the magnetic field. While the incompressibility assumption may be justified
for the fast wind streams (solar minimum) it is much harder to extend its
applicability to the slow wind (solar maximum).
The average isotropy of the magnetic field direction is even harder to realise
in solar wind turbulence, which is inherently anisotropic.
As a corollary, the applicability of ``ideal'' phenomenologies such as 
Irosnikov-Kraichnan is an open question, since the solar wind is anisotropic 
(background magnetic field) and asymmetric (Alfv\'en wave fluxes tend to be 
away from the sun) and compressible~\cite{hnat}.

It might thus be that the coincidence of the magnetic field
magnitude scaling exponents and those of passive scalars is
fortuitous, especially given the fact that the advecting velocity
fields are different. Indeed, one arises from strongly coupled MHD
turbulence, and the other from incompressible fluid turbulence,
and it has been ascertained that the intermittency properties of
passively advected scalars do depend on the velocity
field~\cite{poeta3,vergassola}.  The similarity of
scaling does therefore not imply a similarity of dynamics,
as already noted by~\cite{hnat, ssr05}. We finally offer an
alternative dynamical model that can explain the increased
intermittency of the $|B|$ fluctuations.

It has been shown that the stronger intermittency of the passive
scalar reflects the presence of sharp gradients of the field
within its stochastic fluctuations ~\cite{poeta1,poeta2,poeta3}.
This peculiar behaviour could be the signature of the advected
field being trapped in coherent structures, generated by the
turbulent cascade in the velocity field.  Such large gradients are
characteristic of the slow solar wind (solar maximum) which is
populated with transition zones, that is shearing zones and
shocks. The effect of adding the strong gradients between the
zones is that of increased intermittency which, in turn, makes the
behaviour more similar to that of a passive scalar advected by
ordinary turbulence. If this is the dominant dynamics in an
extended interval of data, then it will dominate the scaling
exponents.

\begin{acknowledgments}

We are grateful to the following people and organizations for the provision of data used 
in this study: R.P. Lepping and and K.W. Ogilvie (both at NASA Goddard Space Flight
Center) for providing  WIND MFI and SWE data, respectively and the ACE MAG instrument 
team and the ACE Science Center for providing the ACE magnetic data. 
We also thank H. Rosenbauer and R. Schwenn (PI's of the plasma instruments onboard Helios 2)
and F. Mariani and N. Ness (PI's of the magnetometer onboard Helios 2) for using their data. 
SCC and BH thank the PPARC and EPSRC for their support.

\end{acknowledgments}



\begin{thebibliography}{32}
\expandafter\ifx\csname natexlab\endcsname\relax\def\natexlab#1{#1}\fi
\expandafter\ifx\csname bibnamefont\endcsname\relax
  \def\bibnamefont#1{#1}\fi
\expandafter\ifx\csname bibfnamefont\endcsname\relax
  \def\bibfnamefont#1{#1}\fi
\expandafter\ifx\csname citenamefont\endcsname\relax
  \def\citenamefont#1{#1}\fi
\expandafter\ifx\csname url\endcsname\relax
  \def\url#1{\texttt{#1}}\fi
\expandafter\ifx\csname urlprefix\endcsname\relax\def\urlprefix{URL }\fi
\providecommand{\bibinfo}[2]{#2}
\providecommand{\eprint}[2][]{\url{#2}}

\bibitem[{\citenamefont{{Tu} and {Marsch}}(1995)}]{tuemarsch}
\bibinfo{author}{\bibfnamefont{C.-Y.} \bibnamefont{{Tu}}} \bibnamefont{and}
  \bibinfo{author}{\bibfnamefont{E.}~\bibnamefont{{Marsch}}},
  \bibinfo{journal}{Space Science Reviews} \textbf{\bibinfo{volume}{73}},
  \bibinfo{pages}{1} (\bibinfo{year}{1995}).

\bibitem[{\citenamefont{{Bruno} and {Carbone}}(2005)}]{living}
\bibinfo{author}{\bibfnamefont{R.}~\bibnamefont{{Bruno}}} \bibnamefont{and}
  \bibinfo{author}{\bibfnamefont{V.}~\bibnamefont{{Carbone}}},
  \bibinfo{journal}{Living Reviews in Solar Physics}
  \textbf{\bibinfo{volume}{2}} (\bibinfo{year}{2005}).

\bibitem[{\citenamefont{{Hnat} et~al.}(2003)\citenamefont{{Hnat}, {Chapman},
  and {Rowlands}}}]{hnatpre}
\bibinfo{author}{\bibfnamefont{B.}~\bibnamefont{{Hnat}}},
  \bibinfo{author}{\bibfnamefont{S.~C.} \bibnamefont{{Chapman}}},
  \bibnamefont{and}
  \bibinfo{author}{\bibfnamefont{G.}~\bibnamefont{{Rowlands}}},
  \bibinfo{journal}{Phys. Rev. E} \textbf{\bibinfo{volume}{67}},
  \bibinfo{pages}{056404} (\bibinfo{year}{2003}).

\bibitem[{\citenamefont{{Kolmogorov}}(1941)}]{k41}
\bibinfo{author}{\bibfnamefont{A.~N.} \bibnamefont{{Kolmogorov}}}, in
  \emph{\bibinfo{booktitle}{Dokl. Akad. Nauk. SSSR}} (\bibinfo{year}{1941}),
  vol.~\bibinfo{volume}{30} of \emph{\bibinfo{series}{Proc. R. Soc. Lond.}},
  pp. \bibinfo{pages}{301--305}, \bibinfo{note}{reprinted in Proc. R. Soc.
  London, A 434, 9--13, 1991}.

\bibitem[{\citenamefont{{Coleman}}(1968)}]{coleman}
\bibinfo{author}{\bibfnamefont{P.}~\bibnamefont{{Coleman}}},
  \bibinfo{journal}{Astrophys. J.} \textbf{\bibinfo{volume}{153}},
  \bibinfo{pages}{371} (\bibinfo{year}{1968}).

\bibitem[{\citenamefont{{Frisch}}(1995)}]{frisch}
\bibinfo{author}{\bibfnamefont{U.}~\bibnamefont{{Frisch}}},
  \emph{\bibinfo{title}{Turbulence. The legacy of A.N. Kolmogorov}}
  (\bibinfo{publisher}{Cambridge: Cambridge University Press, |c1995},
  \bibinfo{year}{1995}).

\bibitem[{\citenamefont{{Burlaga}}(1991)}]{burlaga91}
\bibinfo{author}{\bibfnamefont{L.}~\bibnamefont{{Burlaga}}},
  \bibinfo{journal}{J. Geophys. Res.} \textbf{\bibinfo{volume}{96}},
  \bibinfo{pages}{5847} (\bibinfo{year}{1991}).

\bibitem[{\citenamefont{{Hnat} et~al.}(2005{\natexlab{a}})\citenamefont{{Hnat},
  {Chapman}, and {Rowlands}}}]{hnat}
\bibinfo{author}{\bibfnamefont{B.}~\bibnamefont{{Hnat}}},
  \bibinfo{author}{\bibfnamefont{S.~C.} \bibnamefont{{Chapman}}},
  \bibnamefont{and}
  \bibinfo{author}{\bibfnamefont{G.}~\bibnamefont{{Rowlands}}},
  \bibinfo{journal}{Physical Review Letters} \textbf{\bibinfo{volume}{94}},
  \bibinfo{pages}{204502} (\bibinfo{year}{2005}{\natexlab{a}}).

\bibitem[{\citenamefont{{Carbone} et~al.}(1995)\citenamefont{{Carbone},
  {Veltri}, and {Bruno}}}]{carbone95}
\bibinfo{author}{\bibfnamefont{V.}~\bibnamefont{{Carbone}}},
  \bibinfo{author}{\bibfnamefont{P.}~\bibnamefont{{Veltri}}}, \bibnamefont{and}
  \bibinfo{author}{\bibfnamefont{R.}~\bibnamefont{{Bruno}}},
  \bibinfo{journal}{Physical Review Letters} \textbf{\bibinfo{volume}{75}},
  \bibinfo{pages}{3110} (\bibinfo{year}{1995}).

\bibitem[{\citenamefont{{Marsch} and {Liu}}(1993)}]{marsch&liu93}
\bibinfo{author}{\bibfnamefont{E.}~\bibnamefont{{Marsch}}} \bibnamefont{and}
  \bibinfo{author}{\bibfnamefont{S.}~\bibnamefont{{Liu}}},
  \bibinfo{journal}{Annales Geophysicae} \textbf{\bibinfo{volume}{11}},
  \bibinfo{pages}{227} (\bibinfo{year}{1993}).

\bibitem[{\citenamefont{{Sorriso--Valvo}
  et~al.}(1999)\citenamefont{{Sorriso--Valvo}, {Carbone}, {Veltri},
  {Consolini}, and {Bruno}}}]{sorriso99}
\bibinfo{author}{\bibfnamefont{L.}~\bibnamefont{{Sorriso--Valvo}}},
  \bibinfo{author}{\bibfnamefont{V.}~\bibnamefont{{Carbone}}},
  \bibinfo{author}{\bibfnamefont{P.}~\bibnamefont{{Veltri}}},
  \bibinfo{author}{\bibfnamefont{G.}~\bibnamefont{{Consolini}}},
  \bibnamefont{and} \bibinfo{author}{\bibfnamefont{R.}~\bibnamefont{{Bruno}}},
  \bibinfo{journal}{Geophys. Res. Lett.} \textbf{\bibinfo{volume}{26}},
  \bibinfo{pages}{1801} (\bibinfo{year}{1999}).

\bibitem[{\citenamefont{{Carbone} et~al.}(2004)\citenamefont{{Carbone},
  {Bruno}, {Sorriso-Valvo}, and {Lepreti}}}]{sorriso04}
\bibinfo{author}{\bibfnamefont{V.}~\bibnamefont{{Carbone}}},
  \bibinfo{author}{\bibfnamefont{R.}~\bibnamefont{{Bruno}}},
  \bibinfo{author}{\bibfnamefont{L.}~\bibnamefont{{Sorriso-Valvo}}},
  \bibnamefont{and}
  \bibinfo{author}{\bibfnamefont{F.}~\bibnamefont{{Lepreti}}},
  \bibinfo{journal}{Planet. Spa. Sci.} \textbf{\bibinfo{volume}{52}},
  \bibinfo{pages}{953} (\bibinfo{year}{2004}).

\bibitem[{\citenamefont{{Politano} et~al.}(1998)\citenamefont{{Politano},
  {Pouquet}, and {Carbone}}}]{popoca}
\bibinfo{author}{\bibfnamefont{H.}~\bibnamefont{{Politano}}},
  \bibinfo{author}{\bibfnamefont{A.}~\bibnamefont{{Pouquet}}},
  \bibnamefont{and}
  \bibinfo{author}{\bibfnamefont{V.}~\bibnamefont{{Carbone}}},
  \bibinfo{journal}{Europhysics Letters} \textbf{\bibinfo{volume}{43}},
  \bibinfo{pages}{516} (\bibinfo{year}{1998}).

\bibitem[{\citenamefont{{Sorriso-Valvo}
  et~al.}(2001{\natexlab{a}})\citenamefont{{Sorriso-Valvo}, {Carbone},
  {Veltri}, {Politano}, and {Pouquet}}}]{sorriso00}
\bibinfo{author}{\bibfnamefont{L.}~\bibnamefont{{Sorriso-Valvo}}},
  \bibinfo{author}{\bibfnamefont{V.}~\bibnamefont{{Carbone}}},
  \bibinfo{author}{\bibfnamefont{P.}~\bibnamefont{{Veltri}}},
  \bibinfo{author}{\bibfnamefont{H.}~\bibnamefont{{Politano}}},
  \bibnamefont{and}
  \bibinfo{author}{\bibfnamefont{A.}~\bibnamefont{{Pouquet}}},
  \bibinfo{journal}{Europhysics Letters} \textbf{\bibinfo{volume}{51}},
  \bibinfo{pages}{520} (\bibinfo{year}{2000}).

\bibitem[{\citenamefont{{Sorriso-Valvo}
  et~al.}(2001{\natexlab{b}})\citenamefont{{Sorriso-Valvo}, {Carbone},
  {Giuliani}, {Veltri}, {Bruno}, {Antoni}, and {Martines}}}]{sorriso01}
\bibinfo{author}{\bibfnamefont{L.}~\bibnamefont{{Sorriso-Valvo}}},
  \bibinfo{author}{\bibfnamefont{V.}~\bibnamefont{{Carbone}}},
  \bibinfo{author}{\bibfnamefont{P.}~\bibnamefont{{Giuliani}}},
  \bibinfo{author}{\bibfnamefont{P.}~\bibnamefont{{Veltri}}},
  \bibinfo{author}{\bibfnamefont{R.}~\bibnamefont{{Bruno}}},
  \bibinfo{author}{\bibfnamefont{V.}~\bibnamefont{{Antoni}}}, \bibnamefont{and}
  \bibinfo{author}{\bibfnamefont{E.}~\bibnamefont{{Martines}}},
  \bibinfo{journal}{Planetary and Space Science} \textbf{\bibinfo{volume}{49}},
  \bibinfo{pages}{1193} (\bibinfo{year}{2001}{\natexlab{b}}).

\bibitem[{\citenamefont{{Merrifield}
  et~al.}(2007{\natexlab{c}})\citenamefont{{Merrifield}, {Chapman},
  and {Dendy}}}]{merri}
\bibinfo{author}{\bibfnamefont{J.A.}~\bibnamefont{{Merrifield}}},
  \bibinfo{author}{\bibfnamefont{S.C.}~\bibnamefont{{Chapman}}},
  \bibnamefont{and}
  \bibinfo{author}{\bibfnamefont{R.O.}~\bibnamefont{{Dendy}}},
  \bibinfo{journal}{Physics of Plasmas} \textbf{\bibinfo{volume}{14}},
  \bibinfo{pages}{12301} 
  (\bibinfo{year}{2007}).

\bibitem[{\citenamefont{{Falkovich} et~al.}(2001)\citenamefont{{Falkovich},
  {Gaw{\c e}dzki}, and {Vergassola}}}]{vergassola}
\bibinfo{author}{\bibfnamefont{G.}~\bibnamefont{{Falkovich}}},
  \bibinfo{author}{\bibfnamefont{K.}~\bibnamefont{{Gaw{\c e}dzki}}},
  \bibnamefont{and}
  \bibinfo{author}{\bibfnamefont{M.}~\bibnamefont{{Vergassola}}},
  \bibinfo{journal}{Reviews of Modern Physics} \textbf{\bibinfo{volume}{73}},
  \bibinfo{pages}{913} (\bibinfo{year}{2001}).

\bibitem[{\citenamefont{{Malara} et~al.}(2000)\citenamefont{{Malara},
  {Primavera}, and {Veltri}}}]{malara}
\bibinfo{author}{\bibfnamefont{F.}~\bibnamefont{{Malara}}},
  \bibinfo{author}{\bibfnamefont{L.}~\bibnamefont{{Primavera}}},
  \bibnamefont{and} \bibinfo{author}{\bibfnamefont{P.}~\bibnamefont{{Veltri}}},
  \bibinfo{journal}{Physics of Plasmas} \textbf{\bibinfo{volume}{7}},
  \bibinfo{pages}{2866} (\bibinfo{year}{2000}).

\bibitem[{\citenamefont{{Bavassano} et~al.}(2000)\citenamefont{{Bavassano},
  {Pietropaolo}, and {Bruno}}}]{bavassano2000}
\bibinfo{author}{\bibfnamefont{B.}~\bibnamefont{{Bavassano}}},
  \bibinfo{author}{\bibfnamefont{E.}~\bibnamefont{{Pietropaolo}}},
  \bibnamefont{and} \bibinfo{author}{\bibfnamefont{R.}~\bibnamefont{{Bruno}}},
  \bibinfo{journal}{J. Geophys. Res.}  \textbf{\bibinfo{volume}{105}}, \bibinfo{pages}{15959--15964}
  (\bibinfo{year}{2000}).

\bibitem[{\citenamefont{{Benzi} et~al.}(1993)\citenamefont{{Benzi},
  {Ciliberto}, {Tripiccione}, {Baudet}, {Massaioli}, and {Succi}}}]{ess}
\bibinfo{author}{\bibfnamefont{R.}~\bibnamefont{{Benzi}}},
  \bibinfo{author}{\bibfnamefont{S.}~\bibnamefont{{Ciliberto}}},
  \bibinfo{author}{\bibfnamefont{R.}~\bibnamefont{{Tripiccione}}},
  \bibinfo{author}{\bibfnamefont{C.}~\bibnamefont{{Baudet}}},
  \bibinfo{author}{\bibfnamefont{F.}~\bibnamefont{{Massaioli}}},
  \bibnamefont{and} \bibinfo{author}{\bibfnamefont{S.}~\bibnamefont{{Succi}}},
  \bibinfo{journal}{Phys. Rev. E} \textbf{\bibinfo{volume}{48}},
  \bibinfo{pages}{29} (\bibinfo{year}{1993}).

\bibitem[{\citenamefont{{Benzi} et~al.}(1999)\citenamefont{{Benzi}, {Amati},
  {Casciola}, {Toschi}, and {Piva}}}]{benzi}
\bibinfo{author}{\bibfnamefont{R.}~\bibnamefont{{Benzi}}},
  \bibinfo{author}{\bibfnamefont{G.}~\bibnamefont{{Amati}}},
  \bibinfo{author}{\bibfnamefont{C.~M.} \bibnamefont{{Casciola}}},
  \bibinfo{author}{\bibfnamefont{F.}~\bibnamefont{{Toschi}}}, \bibnamefont{and}
  \bibinfo{author}{\bibfnamefont{R.}~\bibnamefont{{Piva}}},
  \bibinfo{journal}{Physics of Fluids} \textbf{\bibinfo{volume}{11}},
  \bibinfo{pages}{1284} (\bibinfo{year}{1999}).

\bibitem[{\citenamefont{{Hnat} et~al.}(2005{\natexlab{b}})\citenamefont{{Hnat},
  {Chapman}, and {Rowlands}}}]{hnatjgr}
\bibinfo{author}{\bibfnamefont{B.}~\bibnamefont{{Hnat}}},
  \bibinfo{author}{\bibfnamefont{S.~C.} \bibnamefont{{Chapman}}},
  \bibnamefont{and}
  \bibinfo{author}{\bibfnamefont{G.}~\bibnamefont{{Rowlands}}},
  \bibinfo{journal}{Journal of Geophysical Research}
  \textbf{\bibinfo{volume}{110}} (\bibinfo{year}{2005}{\natexlab{b}}).

\bibitem[{\citenamefont{{Bershadskii} and {Sreenivasan}}(2004)}]{papero}
\bibinfo{author}{\bibfnamefont{A.}~\bibnamefont{{Bershadskii}}}
  \bibnamefont{and} \bibinfo{author}{\bibfnamefont{K.~R.}
  \bibnamefont{{Sreenivasan}}}, \bibinfo{journal}{Physical Review Letters}
  \textbf{\bibinfo{volume}{93}}, \bibinfo{pages}{064501}
  (\bibinfo{year}{2004}).

\bibitem[{\citenamefont{{Chapman} et~al.}(2005)\citenamefont{{Chapman}, {Hnat},
  {Rowlands}, and {Watkins}}}]{chapmanNPG05}
\bibinfo{author}{\bibfnamefont{S.~C.} \bibnamefont{{Chapman}}},
  \bibinfo{author}{\bibfnamefont{B.}~\bibnamefont{{Hnat}}},
  \bibinfo{author}{\bibfnamefont{G.}~\bibnamefont{{Rowlands}}},
  \bibnamefont{and} \bibinfo{author}{\bibfnamefont{N.~W.}
  \bibnamefont{{Watkins}}}, \bibinfo{journal}{Nonlinear Processes in
  Geophysics} \textbf{\bibinfo{volume}{12}}, \bibinfo{pages}{767}
  (\bibinfo{year}{2005}).

\bibitem[{\citenamefont{{Hnat} et~al.}(2004)\citenamefont{{Hnat}, {Chapman},
  and {Rowlands}}}]{hnatpop}
\bibinfo{author}{\bibfnamefont{B.}~\bibnamefont{{Hnat}}},
  \bibinfo{author}{\bibfnamefont{S.~C.} \bibnamefont{{Chapman}}},
  \bibnamefont{and}
  \bibinfo{author}{\bibfnamefont{G.}~\bibnamefont{{Rowlands}}},
  \bibinfo{journal}{Physics of Plasmas} \textbf{\bibinfo{volume}{11}},
  \bibinfo{pages}{1326} (\bibinfo{year}{2004}).

\bibitem[{\citenamefont{{Gopalswamy} et~al.}(2004)\citenamefont{{Gopalswamy},
  {Nunes}, {Yashiro}, and {Howard}}}]{Gopalswamy}
\bibinfo{author}{\bibfnamefont{N.}~\bibnamefont{{Gopalswamy}}},
  \bibinfo{author}{\bibfnamefont{S.}~\bibnamefont{{Nunes}}},
  \bibinfo{author}{\bibfnamefont{S.}~\bibnamefont{{Yashiro}}},
  \bibnamefont{and} \bibinfo{author}{\bibfnamefont{R.~A.}
  \bibnamefont{{Howard}}}, \bibinfo{journal}{Advances in Space Research}
  \textbf{\bibinfo{volume}{34}}, \bibinfo{pages}{391} (\bibinfo{year}{2004}).

\bibitem[{\citenamefont{{Matthaeus} et~al.}(2005)\citenamefont{{Matthaeus},
  {Dasso}, {Weygand}, Milano, Smith, and Kivelson}}]{matt05}
\bibinfo{author}{\bibfnamefont{W.~H.} \bibnamefont{{Matthaeus}}},
  \bibinfo{author}{\bibfnamefont{S.}~\bibnamefont{{Dasso}}},
  \bibinfo{author}{\bibfnamefont{J.~M.} \bibnamefont{{Weygand}}},
  \bibinfo{author}{\bibfnamefont{L.~J.} \bibnamefont{Milano}},
  \bibinfo{author}{\bibfnamefont{C.~W.} \bibnamefont{Smith}}, \bibnamefont{and}
  \bibinfo{author}{\bibfnamefont{M.}~\bibnamefont{Kivelson}},
  \bibinfo{journal}{Phys. Rev. lett.} \textbf{\bibinfo{volume}{95}},
  \bibinfo{pages}{231101} (\bibinfo{year}{2005}).

\bibitem[{\citenamefont{{Bruno} et~al.}(2001)\citenamefont{{Bruno}, {Carbone},
  {Veltri}, {Pietropaolo}, and {Bavassano}}}]{bobbybrown2}
\bibinfo{author}{\bibfnamefont{R.}~\bibnamefont{{Bruno}}},
  \bibinfo{author}{\bibfnamefont{V.}~\bibnamefont{{Carbone}}},
  \bibinfo{author}{\bibfnamefont{P.}~\bibnamefont{{Veltri}}},
  \bibinfo{author}{\bibfnamefont{E.}~\bibnamefont{{Pietropaolo}}},
  \bibnamefont{and}
  \bibinfo{author}{\bibfnamefont{B.}~\bibnamefont{{Bavassano}}},
  \bibinfo{journal}{Planetary Space Sci.} \textbf{\bibinfo{volume}{49}},
  \bibinfo{pages}{1201} (\bibinfo{year}{2001}).

\bibitem[{\citenamefont{{Frisch} et~al.}(1999)\citenamefont{{Frisch},
  {Mazzino}, {Noullez}, and {Vergassola}}}]{passive_anz}
\bibinfo{author}{\bibfnamefont{U.}~\bibnamefont{{Frisch}}},
  \bibinfo{author}{\bibfnamefont{A.}~\bibnamefont{{Mazzino}}},
  \bibinfo{author}{\bibfnamefont{A.}~\bibnamefont{{Noullez}}},
  \bibnamefont{and}
  \bibinfo{author}{\bibfnamefont{M.}~\bibnamefont{{Vergassola}}},
  \bibinfo{journal}{Physics of Fluids} \textbf{\bibinfo{volume}{11}},
  \bibinfo{pages}{2178} (\bibinfo{year}{1999}).

\bibitem[{\citenamefont{{Celani} et~al.}(2001)\citenamefont{{Celani},
  {Lanotte}, {Mazzino}, and {Vergassola}}}]{poeta3}
\bibinfo{author}{\bibfnamefont{A.}~\bibnamefont{{Celani}}},
  \bibinfo{author}{\bibfnamefont{A.}~\bibnamefont{{Lanotte}}},
  \bibinfo{author}{\bibfnamefont{A.}~\bibnamefont{{Mazzino}}},
  \bibnamefont{and}
  \bibinfo{author}{\bibfnamefont{M.}~\bibnamefont{{Vergassola}}},
  \bibinfo{journal}{Physics of Fluids} \textbf{\bibinfo{volume}{13}},
  \bibinfo{pages}{1768} (\bibinfo{year}{2001}).

\bibitem[{\citenamefont{{Sorriso-Valvo}
  et~al.}(2006)\citenamefont{{Sorriso-Valvo}, {Carbone}, and {Bruno}}}]{ssr05}
\bibinfo{author}{\bibfnamefont{L.}~\bibnamefont{{Sorriso-Valvo}}},
  \bibinfo{author}{\bibfnamefont{V.}~\bibnamefont{{Carbone}}},
  \bibnamefont{and} \bibinfo{author}{\bibfnamefont{R.}~\bibnamefont{{Bruno}}},
  \bibinfo{journal}{Sp. Sci. Rev.} \textbf{\bibinfo{volume}{121}},
  \bibinfo{pages}{49}  (\bibinfo{year}{2005}).

\bibitem[{\citenamefont{{Sreenivasan}}(1991)}]{poeta1}
\bibinfo{author}{\bibfnamefont{K.~R.} \bibnamefont{{Sreenivasan}}},
  \bibinfo{journal}{Royal Society of London Proceedings Series A}
  \textbf{\bibinfo{volume}{434}}, \bibinfo{pages}{165} (\bibinfo{year}{1991}).

\bibitem[{\citenamefont{{Warhaft}}(2000)}]{poeta2}
\bibinfo{author}{\bibfnamefont{Z.}~\bibnamefont{{Warhaft}}},
  \bibinfo{journal}{Annual Review of Fluid Mechanics}
  \textbf{\bibinfo{volume}{32}}, \bibinfo{pages}{203} (\bibinfo{year}{2000}).

\end{thebibliography}

\clearpage

FIG. 1 CAPTION - The normalized scaling exponents $\zeta_p$ as a function of the moment order
$p$ are reported, along with the linear expected value $p/3$ (full line).
Data refers to the bulk velocity (black circles) and the magnitude of the
magnetic field (white circles), as measured by the Helios 2 satellite in
the inner heliosphere at $0.9$ Astronomical Units during slow wind streams.
Scaling exponents have been obtained through the Extended Self-Similarity
technique. Reported for comparison are the normalized scaling exponents for
longitudinal velocity field (stars) and the temperature field (passive scalar)
in usual fluid flow.

FIG. 2 CAPTION - (Color online) Structure functions of interplanetary magnetic field derived from: 
(a) WIND data for the solar minimum
of 1996 and (b) ACE spacecraft data for the solar maximum of 2000. Solid lines
represent the best linear fit to the points between temporal scales $\sim
10$ minutes and $\sim 10$ hours.

FIG. 3 CAPTION - (Color online) ESS derived from (a) WIND data for the solar minimum of 1996
and (b) ACE spacecraft data for the solar maximum of 2000. Solid lines represent
the best linear fit to the points between temporal scales $\sim 1$ minute and
$\sim 10$ hours.

FIG. 4 CAPTION - (Color online) Scaling exponents from ESS of the magnetic field magnitude fluctuations
for solar minimum (empty circles), solar maximum (filled circles) and the ACE 
interval for $1998$ (triangles).

FIG. 5 CAPTION - Top panel: wind speed profile during 30 days of data recorded by WIND. Bottom panel: magnetic
field intensity profile. The four vertical dashed lines and the three different
symbols identify three different regions where we evaluated the intermittent
character of the fluctuations.

FIG. 6 CAPTION - Flatness factor versus time scale relative to the three different time intervals
shown in the previous Figure.

\clearpage

\clearpage

\clearpage

\clearpage

\clearpage

\clearpage

\end{document}